\documentclass{statsoc}

\usepackage{amsmath,amsfonts,amssymb,mathrsfs}
\usepackage{fullpage}
\usepackage{natbib}
\title{Application of Girsanov Theorem to Particle Filtering
of Discretely Observed Continuous-Time Non-Linear Systems}

\author{Simo S\"arkk\"a}
\address{Helsinki University of Technology \\
  Department of Electrical and Communications Engineering \\
  P.O. Box 9203 \\
  FIN-02015 HUT \\
  Finland}
\email{simo.sarkka@hut.fi (the corresponding author)}

\author{Tommi Sottinen}
\address{
Reykjavik University \\
School of Science and Engineering and School of Business \\
Kringlan 1 \\
IS-103 Reykjavik \\
Iceland}
\email{tommi@ru.is}

\renewcommand{\vec}[1]{\mathbf{#1}}
\newcommand{\set}[1]{\mathcal{#1}}
\newcommand{\alg}[1]{\mathcal{#1}}
\newcommand{\spc}[1]{\mathbb{#1}}
\newcommand{\mea}[1]{\mathbb{#1}}

\renewcommand{\vec}[1]{\mathbf{#1}}
\newcommand{\mat}[1]{\mathit{#1}}

\newcommand{\diff}[0]{\mathrm{d}}

\newcommand{\vectheta}[0]{\mathrm{\boldsymbol{\theta}}}
\newcommand{\vecbeta}[0]{\mathrm{\boldsymbol{\beta}}}
\newcommand{\veceta}[0]{\boldsymbol{\eta}}
\newcommand{\vecmu}[0]{\boldsymbol{\mu}}

\DeclareMathOperator{\diag}{diag}

\DeclareMathOperator{\E}{E}
\DeclareMathOperator{\N}{N}

\newtheorem{algorithm}{Algorithm}[section]

\newtheorem{theorem}{Theorem}[section]

\begin{document}

\begin{abstract}
  This article considers the application of particle filtering to
  continuous-discrete optimal filtering problems, where the system
  model is a stochastic differential equation, and noisy measurements
  of the system are obtained at discrete instances of time. It is
  shown how the Girsanov theorem can be used for evaluating the
  likelihood ratios needed in importance sampling. It is also shown
  how the methodology can be applied to a class of models, where the
  driving noise process is lower in the dimensionality than the state
  and thus the laws of state and noise are not absolutely continuous.
  Rao-Blackwellization of conditionally Gaussian models and unknown
  static parameter models is also considered.
\end{abstract}

\keywords{Girsanov theorem,
particle filtering,
con\-tin\-u\-ous-discrete filtering
}

\section{Introduction}
This article considers the application of sequential importance
sampling based {\em particle filtering} \citep[see,
e.g.][]{Kitagawa:1996,Doucet+Freitas+Gordon:2001} to {\em
  continuous-discrete filtering problems} \citep{Jazwinski:1970},
where the dynamic model is a stochastic differential equation of the
form
\begin{equation}
  \diff \vec{x}(t) = \vec{f}(\vec{x}(t),t) \, \diff t
                 + \mat{L}(t) \, \diff \vecbeta(t),
   \label{eq:sde}
\end{equation}
where $\vec{x}(t) \in \spc{R}^n$ is the state, $\vec{f} : \spc{R}^n
\times \spc{R}_+ \mapsto \spc{R}^n$ is the drift term, $\mat{L}(t) :
\spc{R}^n \times \spc{R}_+ \mapsto \spc{R}^{n \times s}$ is the
dispersion matrix, and $\vecbeta(t) \in \spc{R}^s$ is an
$s$-dimensional Brownian motion with diffusion matrix $\mat{Q}(t)$. It
is assumed that the required conditions
\citep{Karatzas+Shreve:1991,Oksendal:2003} for existence of a strong
solution to the equation are satisfied. In this article, we first
consider the case where the dimensionality of the state is the same as
the dimensionality of the Brownian motion, that is, where $s=n$. We
also extend the results to the singular case where the dimensionality
of the Brownian motion is less than the dimensionality of the state,
that is, where $s < n$.

The likelihood of a measurement $\vec{y}_k$ is modeled by a
probability density, which is a function of the state at time $t_k$:
\begin{equation}
  \vec{y}_k \sim p(\vec{y}_k\,|\,\vec{x}(t_k)).
\end{equation}
The purpose of the {\em Bayesian optimal continuous-discrete filter}
is to compute the posterior distribution (or at least the posterior
mean) of the current state $\vec{x}(t_k)$ given the measurements up to
the current time, that is \citep{Jazwinski:1966,Jazwinski:1970}
\begin{equation}
  p(\vec{x}(t_k)\,|\,\vec{y}_1,\ldots,\vec{y}_k).
\end{equation}
This kind of continuous-discrete filtering models are common in
engineering applications, especially in the fields of navigation,
communication and control
\citep{Bar-Shalom+Li+Kirubarajan:2001,Grewal+Weill+Andrews:2001,Stengel:1994,Van-Trees:1968,Van-Trees:1971}.
In these applications, the dynamic system or a physical phenomenon can
be modeled as a stochastic differential equation, which is observed at
discrete instances of time with certain physical sensors. The purpose
of the filtering or recursive estimation is to infer the state of the
system from the observed noisy measurements.

In this article, novel measure transformation based methods for
con\-tin\-u\-ous-discrete sequential importance resampling \citep[see,
e.g.][]{Gordon+Salmon+Smith:1993,Kitagawa:1996,Pitt+Shepard:1999,Doucet+Freitas+Gordon:2001,Ristic+Arulampalam+Gordon:2004}
are presented. Some of the methods have already been presented in
\citep{Sarkka:2006a,Sarkka:2006b}, but here the methods are
significantly extended. The methods are based on transformations of
probability measures by the Girsanov theorem
\citep{Kallianpur:1980,Karatzas+Shreve:1991,Oksendal:2003}, which is a
theorem from mathematical probability theory. The theorem can be used
for computing likelihood ratios of stochastic processes. It states
that the likelihood ratio of a stochastic process and Brownian motion,
that is, the Radon-Nikodym derivative of the measure of the stochastic
process with respect to the measure of Brownian motion, can be
represented as an exponential martingale which is the solution to a
certain stochastic differential equation.

Measure transformation based approaches are particularly successful in
con\-tin\-u\-ous time filtering \citep{Kallianpur:1980}, but are less
common in con\-tin\-u\-ous-discrete filtering. The general idea of
using the Girsanov theorem in importance sampling of SDEs has been
presented, for example, in \citet{Kloeden+Platen:1999}. Similar ideas
have also been presented by several authors
\citep{Ionides:2004,Crisan+Lyons:1999,Crisan+Gaines+Lyons:1998,Crisan:2003,DelMoral+Miclo:2000}.

\citet{Beskos:2006} considers exact Monte Carlo simulation of a
restricted class of diffusion models, which are observed at discrete
instances of time without any observation error. As shown in the
discussion of the article, the observation errors can be included in
the model. \citet{Fearnhead+Papaspiliopoulos+Roberts:2007} introduces
particles filters for a class of multidimensional diffusion processes,
and the used Monte Carlo sampling methodology is based on the exact
simulation framework of \citet{Beskos:2006}. The difference to the present
methodology is that the methods of \citet{Fearnhead+Papaspiliopoulos+Roberts:2007}
are not based on time-discretization.

\citet{Durham+Gallant:2002} considers simulated maximum
likelihood estimation of parameters of discretely observed stochastic
stochastic differential equations, where all or some of the components
are perfectly observed.  The methods are based on approximating the
transition densities of the processes and modeling the unobserved
sample paths as latent data.
\citet{Golightly+Wilkinson:2006} applies similar methodology to
sequential estimation of state and parameters of stochastic
differential equation models. \citet{Chib+Pitt+Shephard:2004}
considers MCMC based simulation of diffusion driven state space
models. In the article, it is also shown how the methodology can be
applied to particle filtering of such models.


The advantages of the method proposed here over the previously
proposed methods are:
\begin{itemize}
\item Unlike many measure transformation based approaches the
  methodology presented here is not restricted to one-dimensional or
  to SDE models with non-singular dispersion or diffusion matrices.
  The state dimensionality can be higher than the dimensionality of
  the driving Brownian motion, which is equivalent to the case that
  the dispersion/diffusion matrix is singular.

\item The SDE formulation of the likelihood ratio computation allows
  efficient numerical solving of the problem. In particular,
  simulation based approaches \citep{Kloeden+Platen:1999} can be
  applied. Of course, any other numerical methods for SDEs could be
  applied as well.

\item Dispersion (and diffusion) matrices may depend on time, that is,
  the driving process can be time inhomogeneous.

\item The observation errors can be easily modeled and the model
  flexibility is the same as with discrete-time particle filtering.

\item Efficient importance distributions and Rao-Blackwellization can
  be easily used for improving the efficiency of the sampling.
\end{itemize}

\section{Continuous-Discrete Sequential Importance Resampling}

\subsection{Filtering Models}

We shall concentrate into the following four forms of dynamic models:
\begin{enumerate}
\item {\em Non-singular models}, where the dispersion matrices are
  invertible and thus the dimensionality of the process is the same as
  of the driving Brownian motion. The advantage of this kind of
  processes is that their likelihood ratios can be easily evaluated
  using the Girsanov theorem, but the problem is that they are too
  restricted models for many applications.
  
\item {\em Singular models}, where there is non-singular type of
  model, which is embedded inside a deterministic differential
  equation model and thus the joint model is singular because the
  dimensionality of the process is higher than of the driving Brownian
  motion. This kind of models are typical in navigation and stochastic
  control applications, where the deterministic part is typically
  plain integral operator. Because the outer operator is
  deterministic, the likelihood ratios of processes are determined by
  the inner stochastic processes alone and thus importance sampling of
  this kind of process is very similar to the processes of
  non-singular type above.

\item {\em Conditionally Gaussian models}, where a linear stochastic
  differential equation is driven by a model of the non-singular or
  singular type above. This kind of models can be handled such that we
  only sample the inner process and integrate the linear part using
  the Kalman filter. This way we can form a Rao-Blackwellized
  estimate, where the probability density is approximated by a mixture
  of Gaussian distributions.

\item {\em Conjugate static parameter models}, where the model contains
  a static parameter in such conjugate form that certain
  marginalizations can be analytically evaluated. This result in
  particle filter, where only the dynamic state is sampled and the
  sufficient statistics of the static parameter are evaluated at each
  update stage.
\end{enumerate}
\subsection{Non-Singular and Singular Models}
\label{sec:sir1}
Assume that the filtering model is of the form
\begin{equation}
\begin{split}
   \diff \vec{x} &= \vec{f}(\vec{x},t) \, \diff t
   + \mat{L}(t) \, \diff \vecbeta \\
   \vec{y}_k &\sim
   p(\vec{y}_k\,|\,\vec{x}(t_k)),
\end{split}
\label{eq:abs_cdmodel}
\end{equation}
where $\vecbeta(t)$ is a Brownian motion with positive definite
diffusion matrix $\mat{Q}(t)$, $\mat{L}(t)$ is an invertible matrix
for all $t \ge 0$ and the initial conditions are $\vec{x}(0) \sim
p(\vec{x}(0))$. Further assume that we have constructed an importance
process $\vec{s}(t)$, which is defined by the SDE
\begin{equation}
   \diff \vec{s} = \vec{g}(\vec{s},t) \, \diff t
   + \mat{B}(t) \, \diff \vecbeta,
\label{eq:abs_imp}
\end{equation}
and which has a probability law that is a rough approximation to the
filtering (or smoothing) distribution of the model
\eqref{eq:abs_cdmodel}, at least at the measurement times. The matrix
$\mat{B}(t)$ is also assumed to be invertible for all $t \ge 0$. Note
that at this point we do not want to restrict the matrix $\mat{B}(t)$
to be the same as $\mat{L}(t)$, because this allows usage of greater
class of importance processes as shall be seen later in this article.

Now it is possible to generate a set of importance samples from the
conditioned (i.e., filtered) process $\vec{x}(t)$, which is
conditional to the measurements $\vec{y}_{1:k}$ using $\vec{s}(t)$ as
the importance process. The motivation of this is that because the
process $\vec{s}(t)$ already is an approximation to the optimal
result, using it as the importance process is likely to produce a less
degenerate particle set and thus more accurate presentation of the
filtering distribution.

Because the matrices $\mat{L}(t)$ and $\mat{B}(t)$ are invertible, the
probability measures of $\vec{x}$ and $\vec{s}$ are absolutely
continuous with respect to the probability measure of the driving
Brownian motion $\vecbeta(t)$ and it is possible to compute likelihood
ratio between the target and importance processes by applying the
Girsanov theorem. The explicit expression and derivation of this
likelihood ratio is given in Theorem \ref{the:trans1} of Appendix
\ref{app:lh}.

The SIR algorithm recursion starts by drawing samples $\{
\vec{x}^{(i)}_{0} \}$ from the initial distribution and setting
$w^{(i)}_{0} = 1/N$, where $N$ is the number of Monte Carlo samples.
The con\-tin\-u\-ous-discrete SIR filter algorithm then proceeds as
follows:
\begin{algorithm}[CD-SIR I] \label{alg:cdsir1}
 Given the importance process $\vec{s}(t)$, a weighted set of samples
  $\{ \vec{x}^{(i)}_{k-1}, w^{(i)}_{k-1} \}$ and the new measurement
  $\vec{y}_k$, a single step of {\em con\-tin\-u\-ous-discrete sequential
  importance resampling} can be performed as follows:
\begin{enumerate}
\item Simulate $N$ realizations of the importance processes
  \begin{align}
      \diff \vec{s}^{(i)} &= \vec{g}(\vec{s}^{(i)},t) \, \diff t
      + \mat{B}(t) \, \diff \vecbeta^{(i)}(t), \quad
        &\vec{s}^{(i)}(t_{k-1}) = \vec{x}^{(i)}_{k-1} \nonumber \\
      \diff \vec{s}^{*(i)}(t) &= \mat{L}(t) \, \mat{B}^{-1}(t) \,
      \diff \vec{s}^{(i)}(t), \quad
        &\vec{s}^{*(i)}(t_{k-1}) = \vec{x}^{(i)}_{k-1}, \nonumber
  \end{align}
  from $t=t_{k-1}$ to $t=t_k$, where $\vecbeta^{(i)}(t)$ are
  independent Brownian motions, and $i=1,\ldots,N$.

\item At the same time, simulate the log-likelihood ratios
  \begin{equation}
  \begin{split}
    \diff \Lambda^{(i)} &= 
    \big\{\vec{f}(\vec{s}^{*(i)}(t),t)
  - \mat{L}(t) \, \mat{B}^{-1}(t) \,
    \vec{g}(\vec{s}^{(i)}(t),t)\big\}^T \\
  &\quad \times \mat{L}^{-T}(t) \,
    \mat{Q}^{-1}(t) \, \diff \vecbeta^{(i)}(t)
  \\
  & - \frac{1}{2}
    \big\{\vec{f}(\vec{s}^{*(i)}(t),t)
  - \mat{L}(t) \, \mat{B}^{-1}(t) \,
  \vec{g}(\vec{s}^{(i)}(t),t)\big\}^T \\
  &\quad \times
     \big\{ \mat{L}(t) \, \mat{Q}(t) \, \mat{L}^T(t) \big\}^{-1} \\
  &\quad \times 
   \big\{\vec{f}(\vec{s}^{*(i)}(t),t)
  - \mat{L}(t) \, \mat{B}^{-1}(t) \, \vec{g}(\vec{s}^{(i)}(t),t)\big\} \,
    \diff t, \\
  \Lambda^{(i)}(t_{k-1}) &= 0,
  \end{split}
  \nonumber
  \end{equation}
  from $t=t_{k-1}$ to $t=t_k$ and set
  \begin{equation}
  \begin{split}
    \vec{x}^{(i)}_k &= \vec{s}^{*(i)}(t_k) \\
          Z^{(i)}_k &= \exp\left\{ \Lambda^{(i)}(t_k) \right\}.
  \end{split}
  \nonumber
  \end{equation}
  Note that the realizations of Brownian motions must be the same as
  in simulation of the importance processes.

\item For each $i$ compute
\begin{equation}
\begin{split}
  w^{(i)}_{k} &= w^{(i)}_{k-1} \, Z^{(i)}_k \, p(\vec{y}_k\,|\,\vec{x}^{(i)}_k),
\end{split}
\nonumber
\end{equation}
and re-normalize the weights to sum to unity.

\item If the effective number of particles is too low, perform
  resampling.
\end{enumerate}
\end{algorithm}
Some practical points about the implementation:
\begin{itemize}
\item The importance process $\vec{s}(t)$ required by the algorithm
  can be obtained by using, for example, the extended Kalman filter
  (EKF). An example of this approach is given in Section
  \ref{sec:simu} of this article.

\item The simulation of the importance processes and likelihood ratios
  above can be performed using any of the well known numerical methods
  for simulation of stochastic differential equations
  \citep{Kloeden+Platen:1999}. In this article we have used the simple
  Euler-Maruyama method, which can be considered as a stochastic
  version of the Euler integration for non-stochastic differential
  equations.
\end{itemize}
The class \eqref{eq:abs_cdmodel} is actually very restricted class of
dynamic models, where it is required that the probability law of the
state is absolutely continuous with respect to the law of the driving
Brownian motion. This kind of models are common in mathematical
treatment of stochastic differential equations and such models can be
found, for example, in mathematical finance \citep[see, e.g.,
][]{Karatzas+Shreve:1991,Oksendal:2003}. However, most of the models
used in navigation and telecommunications applications do not fit into
this class, and for this reason the results need to be extended.

It is also possible to construct a similar SIR algorithm for more
general models, where there is an absolutely continuous type of model,
which is {\em embedded} inside a {\em deterministic} differential
equation model.  This kind of models are typical in navigation,
communication and stochastic control applications
\citep{Bar-Shalom+Li+Kirubarajan:2001,Grewal+Weill+Andrews:2001,Stengel:1994,Van-Trees:1968,Van-Trees:1971},
where the deterministic part is typically a plain integral operator.
Because the outer operator is deterministic, the likelihood ratios of
processes are determined by the inner stochastic processes alone and
thus importance sampling of this kind of process is very similar to
sampling of the processes considered above.

Assume that the model is of the form
\begin{equation}
\begin{split}
  \frac{\diff \vec{x}_1}{\diff t} &= \vec{f}_1(\vec{x}_1,\vec{x}_2,t),
  \\
  \diff \vec{x}_2 &= \vec{f}_2(\vec{x}_1,\vec{x}_2,t) \, \diff t
   + \mat{L}(t) \, \diff \vecbeta \\
   \vec{y}_k &\sim
    p(\vec{y}_k\,|\,\vec{x}_1(t_k),\vec{x}_2(t_k)),
\end{split}
\label{eq:lh2model}
\end{equation}
where $\vec{f}_1(\cdot)$ and $\vec{f}_2(\cdot)$ are deterministic
functions, $\vecbeta(t)$ is a Brownian motion, $\mat{L}(t)$ is
invertible matrix and the initial conditions are
$\vec{x}_1(0),\vec{x}_2(0) \sim p(\vec{x}_1(0),\vec{x}_2(0))$. Note
that because the dimensionality of Brownian motion is less than of the
joint state $(\vec{x}_1~\vec{x}_2)^T$ it is not possible to compute
the likelihood ratio between the process and Brownian motion by the
Girsanov theorem directly.

However, it turns out that if the importance process for
$(\vec{x}_1~\vec{x}_2)^T$ is formed as follows
\begin{equation}
\begin{split}
  \frac{\diff \vec{s}_1}{\diff t} &= \vec{f}_1(\vec{s}_1,\vec{s}_2,t) \\
  \diff \vec{s}_2 &= \vec{g}_2(\vec{s}_1,\vec{s}_2,t) \, \diff t
  + \mat{B}(t) \, \diff \vecbeta,
\end{split}
\end{equation}
then the importance weights can be computed in exactly the same way as
when forming importance sample of $\vec{x}_2(t)$ using $\vec{s}_2(t)$
as the importance process. 

The likelihood ratio expression is given in Theorem \ref{the:trans2}
of Appendix \ref{app:lh}. The SIR algorithm is started by first
drawing samples from the initial distribution and then for each
measurement, the following steps are performed:
\begin{algorithm}[CD-SIR II] \label{alg:cdsir2} Given the importance
  process $\vec{s}_1(t),\vec{s}_2(t)$, a weighted set of samples $\{
  \vec{x}^{(i)}_{1,k-1}, \vec{x}^{(i)}_{2,k-1}, w^{(i)}_{k-1} \}$ and
  the new measurement $\vec{y}_k$, a single step of {\em
    con\-tin\-u\-ous-discrete sequential importance resampling} can be 
  performed as follows:
\begin{enumerate}
\item Simulate $N$ realizations of the importance processes
  \begin{align}
    \frac{\diff \vec{s}^{(i)}_1}{\diff t} &= 
      \vec{f}_1(\vec{s}^{(i)}_1,\vec{s}^{(i)}_2,t),
    \quad &\vec{s}^{(i)}_1(t_{k-1}) = \vec{x}^{(i)}_{1,k-1} \nonumber\\
    \diff \vec{s}^{(i)}_2 &= \vec{g}_2(\vec{s}^{(i)}_1,\vec{s}^{(i)}_2,t) \, \diff t
  + \mat{B}(t) \, \diff \vecbeta^{(i)}(t),
     \quad &\vec{s}^{(i)}_2(t_{k-1}) = \vec{x}^{(i)}_{2,k-1} \nonumber\\
  \frac{\diff \vec{s}^{*(i)}_1}{\diff t} &= 
    \vec{f}_1(\vec{s}^{*(i)}_1,\vec{s}^{*(i)}_2,t),
    \quad &\vec{s}^{*(i)}_1(t_{k-1}) = \vec{x}^{(i)}_{1,k-1} \nonumber\\
  \diff \vec{s}^{*(i)}_2 &= \mat{L}(t) \, \mat{B}^{-1}(t) \, \diff \vec{s}_2,
  \quad &\vec{s}^{*(i)}_2(t_{k-1}) = \vec{x}^{(i)}_{2,k-1}, \nonumber
  \end{align}
\item Simulate the log-likelihood ratios (using the same Brownian
  motion realizations as above)
  \begin{equation}
  \begin{split}
    \diff \Lambda^{(i)} &= 
    \big\{\vec{f}_2(\vec{s}^{*(i)}_1(t),\vec{s}^{*(i)}_2(t),t)
  - \mat{L}(t) \, \mat{B}^{-1}(t) \,
    \vec{g}_2(\vec{s}_1^{(i)}(t),\vec{s}_2^{(i)}(t),t)\big\}^T \\
  &\quad 
    \times \mat{L}^{-T}(t) \, \mat{Q}^{-1}(t) \, \diff \vecbeta^{(i)}(t) \\
  &- \frac{1}{2}
    \big\{\vec{f}_2(\vec{s}^{*(i)}_1(t),\vec{s}^{*(i)}_2(t),t)
  - \mat{L}(t) \, \mat{B}^{-1}(t) \, \vec{g}_2(\vec{s}_1^{(i)}(t),\vec{s}^{(i)}_2(t),t)\big\}^T \\
  &\quad \times \big\{ \mat{L}(t) \, \mat{Q}(t) \, \mat{L}^T(t)
  \big\}^{-1} \\
  &\quad 
   \times \big\{\vec{f}_2(\vec{s}^{*(i)}_1(t),\vec{s}^{*(i)}_2(t),t)
  - \mat{L}(t) \, \mat{B}^{-1}(t) \, \vec{g}_2(\vec{s}^{(i)}_1(t),\vec{s}^{(i)}_2(t),t)\big\} \,
    \diff t, \\
  & \Lambda^{(i)}(t_{k-1}) = 0,
  \end{split}
  \nonumber
  \end{equation}
  from $t=t_{k-1}$ to $t=t_k$ and set
  \begin{equation}
  \begin{split}
    \vec{x}^{(i)}_{1,k} &= \vec{s}^{*(i)}_1(t_k) \\
    \vec{x}^{(i)}_{2,k} &= \vec{s}^{*(i)}_2(t_k) \\
          Z^{(i)}_k &= \exp\left\{ \Lambda^{(i)}(t_k) \right\}.
  \end{split}
  \end{equation}

\item For each $i$ compute
\begin{equation}
\begin{split}
  w^{(i)}_{k} &= w^{(i)}_{k-1} \, Z^{(i)}_k \,
    p(\vec{y}_k\,|\,\vec{x}^{(i)}_{1,k},\vec{x}^{(i)}_{2,k}),
\end{split}
\end{equation}
and re-normalize the weights to sum to unity.

\item If the effective number of particles is too low, perform
  resampling.
\end{enumerate}
\end{algorithm}
The importance process $\vec{s}(t)$ required by the algorithm can be
obtained by using, for example, continuous-discrete EKF and
then extracting the estimate of the inner process $\vec{s}_2(t)$ from
the joint estimate.

\subsection{Rao-Blackwellization of Conditionally Gaussian Models}
\label{sec:cdrbsir1}
Now we shall consider dynamic models, where a {\em linear} stochastic
differential equation is driven by a singular or non-singular
model considered in the previous section. This kind of models can be
handled such that only the inner process is sampled and the linear
part is integrated out using the con\-tin\-u\-ous-discrete Kalman
filter. Then it is possible to form a Rao-Blackwellized estimate,
where the probability density is approximated by a mixture of Gaussian
distributions.  The measurement model is assumed to be of the same
form as in previous sections, but linear with respect to the state
variables corresponding to the linear part of the dynamic process.

Dynamic models with conditionally Gaussian parts arise, for example,
when the measurement noise correlations are modeled with state
augmentation \citep[see, e.g., ][]{Gelb:1974}. Actually, in this case,
the direct application of particle filter without Rao-Blackwellization
would be impossible because the measurement model is formally
singular. However, the Rao-Blackwellized filter can be easily applied
to this kind of models.

Assume that the dynamic model is of the form
\begin{equation}
\begin{split}
  \diff \vec{x}_1 &= 
  \vec{F}(\vec{x}_2,\vec{x}_3,t) \, \vec{x}_1 \, \diff t
  + \vec{f}_1(\vec{x}_2,\vec{x}_3,t) \, \diff t 
  + \mat{V}(\vec{x}_2,\vec{x}_3,t) \, \diff \veceta  \\
  \frac{\diff \vec{x}_2}{\diff t} &= \vec{f}_2(\vec{x}_2,\vec{x}_3,t) \\
  \diff \vec{x}_3 &= \vec{f}_3(\vec{x}_2,\vec{x}_3,t) \, \diff t
  + \mat{L}(t) \, \diff \vecbeta,
\end{split}
\label{eq:cdrb_model1}
\end{equation}
where $\veceta$ and $\vecbeta$ are independent Brownian motions with
diffusion matrices $\mat{Q}_{\eta}(t)$ and $\mat{Q}_{\beta}(t)$,
respectively. Also assume that the initial conditions are given as:
\begin{equation}
\begin{split}
  \vec{x}_1(0) &\sim \N(\vec{m}_0,\vec{P}_0) \\
  \vec{x}_2(0),\vec{x}_3(0) &\sim p(\vec{x}_2(0),\vec{x}_3(0)),
\end{split}
\end{equation}
and the initial conditions of $\vec{x}_1(0)$ are independent from those
of $\vec{x}_2(0)$ and $\vec{x}_3(0)$. 

In this case an importance process can be formed as
\begin{equation}
\begin{split}
  \diff \vec{s}_1 &= 
  \vec{F}(\vec{s}_2,\vec{s}_3,t) \, \vec{s}_1 \, \diff t
  + \vec{f}_1(\vec{s}_2,\vec{s}_3,t) \, \diff t + \mat{V}(\vec{s}_2,\vec{s}_3,t) \, \diff \veceta, 
  \\
  \frac{\diff \vec{s}_2}{\diff t} &= \vec{f}_2(\vec{s}_2,\vec{s}_3,t)
  \\
  \diff \vec{s}_3 &= \vec{g}_3(\vec{s}_2,\vec{s}_3,t) \, \diff t
  + \mat{B}(t) \, \diff \vecbeta,
\end{split}
\label{eq:cdrb_imp1} 
\end{equation}
with the same initial conditions. In both the original and importance
processes, conditionally to the filtration of the second Brownian
motion $\alg{F}_t = \sigma(\vecbeta(s), 0 \le s \le t)$ and to initial
conditions, the law of the first equation is determined by the mean
and covariance of the Gaussian process, which is driven by the process
$\veceta(t)$. Thus, conditionally to $\vec{x}_2$ and $\vec{x}_3$ the
process $\vec{x}_1(t)$ is Gaussian for all $t$. The same applies to
the importance process.

Now it is possible to integrate out the Gaussian parts of both 
processes. This procedure results in the following marginalized
equations for the original process:
\begin{align}
  \frac{\diff \vec{m}_x(t)}{\diff t} &= \mat{F}(\vec{x}_2,\vec{x}_3,t) \, \vec{m}_x(t)
  + \vec{f}_1(\vec{x}_2,\vec{x}_3,t)\nonumber \\
  \frac{\diff \vec{P}_x(t)}{\diff t} &= \mat{F}(\vec{x}_2,\vec{x}_3,t) \, \mat{P}_x(t)
  + \mat{P}_x(t) \, \mat{F}^T(\vec{x}_2,\vec{x}_3,t) \nonumber \\
  &+ \mat{V}(\vec{x}_2,\vec{x}_3,t) \, \vec{Q}_{\eta}(t) \,
     \mat{V}^T(\vec{x}_2,\vec{x}_3,t) \label{eq:cdrb_model2} \\
  \frac{\diff \vec{x}_2}{\diff t} &= \vec{f}_2(\vec{x}_2,\vec{x}_3,t) \nonumber \\
  \diff \vec{x}_3 &= \vec{f}_3(\vec{x}_2,\vec{x}_3,t) \, \diff t
  + \mat{L}(t) \, \diff \vecbeta, \nonumber
\end{align}
where $\vec{m}_x(t)$ and $\mat{P}_x(t)$ are the mean and covariance of
the Gaussian process. For the importance process we get similarly:
\begin{align}
  \frac{\diff \vec{m}_s(t)}{\diff t} &= \mat{F}(\vec{s}_2,\vec{s}_3,t) \, \vec{m}_s(t)
  + \vec{f}_1(\vec{s}_2,\vec{s}_3,t) \nonumber \\
  \frac{\diff \vec{P}_s(t)}{\diff t} &= \mat{F}(\vec{s}_2,\vec{s}_3,t) \, \mat{P}_s(t)
  + \mat{P}_s(t) \, \mat{F}^T(\vec{s}_2,\vec{s}_3,t) \nonumber \\
  &+ \mat{V}(\vec{s}_2,\vec{s}_3,t) \, \vec{Q}_{\eta}(t) \,
     \mat{V}^T(\vec{s}_2,\vec{s}_3,t) \label{eq:cdrb_imp2} \\
  \frac{\diff \vec{s}_2}{\diff t} &= \vec{f}_2(\vec{s}_2,\vec{s}_3,t)
  \nonumber \\
  \diff \vec{s}_3 &= \vec{g}_3(\vec{s}_2,\vec{s}_3,t) \, \diff t
  + \mat{B}(t) \, \diff \vecbeta, \nonumber
\end{align}
The models \eqref{eq:cdrb_model2} and \eqref{eq:cdrb_imp2} have now the
form, where the Algorithm \ref{alg:cdsir2} can be used. The
importance sampling now results in the set of weighted samples
\begin{equation}
  \{ w^{(i)}, \vec{m}^{(i)}, \mat{P}^{(i)},
     \vec{x}_2^{(i)}, \vec{x}_3^{(i)} \},
\end{equation}
such that the probability density of the state $\vec{x}(t) =
(\vec{x}_1(t),\vec{x}_2(t),\vec{x}_3(t))$ is approximately given as
\begin{equation}
\begin{split}
  &p(\vec{x}_1(t),\vec{x}_2(t),\vec{x}_3(t)) \\
  &\approx
  \sum_i w^{(i)} \,
  \N(\vec{x}_1(t)\,|\,\vec{m}^{(i)}, \mat{P}^{(i)}) \,
  \delta(\vec{x}_2(t) - \vec{x}_2^{(i)}) \,
  \delta(\vec{x}_3(t) - \vec{x}_3^{(i)}).
\end{split}
\end{equation}
where $\delta(\cdot)$ is the Dirac delta function. If the measurement
model is of the form
%
%
\begin{equation}
  p(\vec{y}_k\,|\,\vec{x}(t_k))
  = \N\left(\vec{y}_k\,|\,\mat{H}_k\left(\vec{x}_2(t_k),\vec{x}_3(t_k)\right) \, 
    \vec{x}_1(t_k),
     \mat{R}_k\left(\vec{x}_2(t_k),\vec{x}_3(t_k)\right)\right),
\end{equation}
then conditionally to $\vec{x}_2(t_k),\vec{x}_3(t_k)$ also the
measurement model is linear Gaussian and the Kalman filter update
equations can be applied. The resulting algorithm is the following:
\begin{algorithm}[CDRB-SIR I] \label{alg:cdrbsir1} Given the importance
  process, a set of importance samples $\{ \vec{x}^{(i)}_{2,k-1},
  \vec{x}^{(i)}_{3,k-1}, \vec{m}^{(i)}_{k-1}, \mat{P}^{(i)}_{k-1},
  w^{(i)}_{k-1} ~:~i=1,\ldots,N \}$ and the measurement $\vec{y}_k$, a
  single step of {\em conditionally Gaussian continuous-discrete
    Rao-Blackwellized SIR} is the following:
\begin{enumerate}
\item Simulate $N$ realizations of the importance process
  \begin{equation}
  \begin{split}
    \frac{\diff \vec{m}_s^{(i)}}{\diff t} &=
    \mat{F}(\vec{s}^{(i)}_2(t),\vec{s}^{(i)}_3,t) \, \vec{m}_s^{(i)}(t)
    + \vec{f}_1(\vec{s}^{(i)}_2,\vec{s}^{(i)}_3,t) \\
    \frac{\diff \mat{P}_s^{(i)}}{\diff t} &=
    \mat{F}(\vec{s}^{(i)}_2,\vec{s}^{(i)}_3,t) \, \mat{P}_s^{(i)}(t)
    + \mat{P}_s^{(i)}(t) \, \mat{F}^T(\vec{s}^{(i)}_2,\vec{s}^{(i)}_3,t)
    \\
     &+ \vec{V}(\vec{s}^{(i)}_2,\vec{s}^{(i)}_3,t) \, \vec{Q}_{\eta}(t) \,
     \mat{V}^T(\vec{s}^{(i)}_2,\vec{s}^{(i)}_3,t) \\
    \frac{\diff \vec{s}^{(i)}_2}{\diff t} &=
    \vec{f}_2(\vec{s}^{(i)}_2,\vec{s}^{(i)}_3,t) \\
    \diff \vec{s}^{(i)}_3 &= \vec{g}_3(\vec{s}^{(i)}_2,\vec{s}^{(i)}_3,t)
      \, \diff t + \mat{B}(t) \, \diff \vecbeta^{(i)}, 
  \end{split}
  \end{equation}
  with initial conditions
  \begin{equation}
  \begin{split}
    \vec{m}_s^{(i)}(t_{k-1}) &=
      \vec{m}^{(i)}_{k-1} \\
    \mat{P}_s^{(i)}(t_{k-1}) &=
      \mat{P}^{(i)}_{k-1} \\
    \vec{s}^{(i)}_2(t_{k-1}) &=
      \vec{x}^{(i)}_{2,k-1} \\
    \vec{s}^{(i)}_3(t_{k-1}) &=
      \vec{x}^{(i)}_{3,k-1},
  \end{split}
  \end{equation}
\item Simulate the scaled importance process
  \begin{equation}
  \begin{split}
    \frac{\diff \vec{m}_s^{*(i)}}{\diff t} &=
    \mat{F}(\vec{s}^{*(i)}_2(t),\vec{s}^{*(i)}_3,t) \, \vec{m}_s^{*(i)}(t)
    + \vec{f}_1(\vec{s}^{*(i)}_2,\vec{s}^{*(i)}_3,t) \\
    \frac{\diff \mat{P}_s^{*(i)}}{\diff t} &=
    \mat{F}(\vec{s}^{*(i)}_2,\vec{s}^{*(i)}_3,t) \, \mat{P}_s^{*(i)}(t)
    + \mat{P}_s^{*(i)}(t) \, \mat{F}^T(\vec{s}^{*(i)}_2,\vec{s}^{*(i)}_3,t)
    \\
     &+ \vec{V}(\vec{s}^{*(i)}_2,\vec{s}^{*(i)}_3,t) \, \vec{Q}_{\eta}(t) \,
     \mat{V}^T(\vec{s}^{*(i)}_2,\vec{s}^{*(i)}_3,t) \\    
    \frac{\diff \vec{s}^{*(i)}_2}{\diff t} &=
    \vec{f}_2(\vec{s}^{*(i)}_2,\vec{s}^{*(i)}_3,t) \\
    \diff \vec{s}^{*(i)}_3 &= \mat{L}(t) \, \mat{B}^{-1}(t) \, \diff \vec{s}_3,
  \end{split}
  \end{equation}
  with the same initial conditions from $t=t_{k-1}$ to $t=t_k$ and set
  \begin{equation}
  \begin{split}
      \vec{m}^{-(i)}_k &= \vec{m}_s^{*(i)}(t_k) \\
      \mat{P}^{-(i)}_k &= \mat{P}_s^{*(i)}(t_k) \\
    \vec{x}^{(i)}_{2,k} &= \vec{s}^{*(i)}_2(t_k) \\
    \vec{x}^{(i)}_{3,k} &= \vec{s}^{*(i)}_3(t_k).
  \end{split}
  \end{equation}

\item Simulate the log-likelihood ratios (again, using the same Brownian
  motion realizations as in importance process)
  \begin{equation}
  \begin{split}
    \diff \Lambda^{(i)} &= 
    \big\{\vec{f}_3(\vec{s}^{*(i)}_2(t),\vec{s}^{*(i)}_3(t),t)
  - \mat{L}(t) \, \mat{B}^{-1}(t) \,
    \vec{g}_3(\vec{s}_2^{(i)}(t),\vec{s}_3^{(i)}(t),t)\big\}^T \\
  &\quad \times \mat{L}^{-T}(t) \, \mat{Q}_{\beta}^{-1}(t) \, \diff \vecbeta^{(i)}(t)
  \\
  &- \frac{1}{2}
    \big\{\vec{f}_3(\vec{s}^{*(i)}_2(t),\vec{s}^{*(i)}_3(t),t)
  - \mat{L}(t) \, \mat{B}^{-1}(t) \, \vec{g}_3(\vec{s}_2^{(i)}(t),\vec{s}^{(i)}_3(t),t)\big\}^T
  \\
  &\quad \times \big\{ \mat{L}(t) \, \mat{Q}_{\beta}(t) \, \mat{L}^T(t)
  \big\}^{-1} \\
  &\quad \times 
   \big\{\vec{f}_3(\vec{s}^{*(i)}_2(t),\vec{s}^{*(i)}_3(t),t)
  - \mat{L}(t) \, \mat{B}^{-1}(t) \, \vec{g}_3(\vec{s}^{(i)}_2(t),\vec{s}^{(i)}_3(t),t)\big\} \,
    \diff t, \\
  & \Lambda^{(i)}(t_{k-1}) = 0,
  \end{split}
  \end{equation}
and set
  \begin{equation}
          Z^{(i)}_k = \exp\left\{ \Lambda^{(i)}(t_k) \right\}
  \end{equation}

\item For each $i$ perform the Kalman filter update
  \begin{equation}
  \begin{split}
   \vecmu^{(i)}_k &=
     \mat{H}_k(\vec{x}^{(i)}_{2,k},\vec{x}^{(i)}_{3,k}) \,
      \vec{m}^{-(i)}_k \\
   \mat{S}_k^{(i)} &=
     \mat{H}_k(\vec{x}^{(i)}_{2,k},\vec{x}^{(i)}_{3,k}) \,
    \mat{P}^{-(i)}_k \,
     \mat{H}_k^T(\vec{x}^{(i)}_{2,k},\vec{x}^{(i)}_{3,k}) +
     \mat{R}_k(\vec{x}^{(i)}_{2,k},\vec{x}^{(i)}_{3,k}) \\
   \mat{K}^{(i)}_k &=
     \mat{P}^{-(i)}_k \,
     \mat{H}_k^T(\vec{x}^{(i)}_{2,k},\vec{x}^{(i)}_{3,k}) \,
     \{\mat{S}^{(i)}_{k}\}^{-1} \\
  \vec{m}^{(i)}_k &= \vec{m}^{-(i)}_k
  + \mat{K}^{(i)}_k \, (\vec{y}_k - \vecmu^{(i)}_k) \\
  \vec{P}^{(i)}_k &= \vec{P}^{-(i)}_k
  - \mat{K}^{(i)}_k \, \mat{S}^{(i)}_k \, \{\mat{K}^{(i)}_k\}^T,
  \end{split}
  \end{equation}
compute the importance weight
\begin{align}
  w^{(i)}_k &= w^{(i)}_{k-1} \, Z^{(i)}_k 
  \, \N(\vec{y}_k\,|\,\vecmu^{(i)}_k,\mat{S}^{(i)}_k), 
\end{align}
and re-normalize the weights to sum to unity.

\item If the effective number of particles is too low,
  perform resampling.
\end{enumerate}
\end{algorithm}
The importance process can be formed, for example, by computing a
joint Gaussian approximation by EKF and then extracting only
the estimates corresponding to the innermost process. Note that the
Rao-Blackwellization procedure can be often performed approximately,
even when the model is not completely Gaussian. The Kalman filter
steps can be replaced with the corresponding steps of EKF, when
the model is slightly non-linear. This approach has been successfully
applied in the context of multiple target tracking in article
\citep{Sarkka+Vehtari+Lampinen:2007a}.

\subsection{Rao-Blackwellization of Models with Static Parameters}
\label{sec:cdrbsir2}
Analogously to the discrete time case presented in
\cite{Storvik:2002}, the procedure of Rao-Blackwellization can often
be applied to models with unknown static parameters. If the posterior
distribution of the unknown static parameters $\vectheta$ depends only
on a suitable set of sufficient statistics $\set{T}_k =
\set{T}_k(\vec{x}(t_1),\ldots,\vec{x}(t_k),\vec{y}_{1:k})$, the
parameter can be marginalized out analytically and only the state
needs to be sampled.

This kind of models arise, for example, when the measurement noise
variance or some other other parameters of the measurement model are
unknown. Two models of this kind are presented in Section
\ref{sec:simu}.

Assume that the model is of the form
\begin{equation}
\begin{split}
  \diff \vec{x} &= \vec{f}(\vec{x},t,\vectheta) \, \diff t
   + \mat{L}(t,\vectheta) \, \diff \vecbeta \\
   \vec{y}_k &\sim   
    p(\vec{y}_k\,|\,\vec{x}(t_k),\vectheta),
\end{split}
\end{equation}
where $\vectheta$ is an unknown static parameter. Also assume that
$\vec{f}(\cdot)$ and $\mat{L}(\cdot)$ are of such form that the model
is either non-singular or singular model considered in Sections
\ref{sec:sir1}.

Now assume that the prior distribution of $\vectheta$ has some finite
dimensional sufficient statistics $\set{T}_0$:
\begin{equation}
  p(\vectheta) = p(\vectheta\,|\,\set{T}_0),
\end{equation}
also assume that conditional posterior distribution of $\vectheta$ has
sufficient statistics $\set{T}_k =
\set{T}_k(\vec{x}(t_1),\ldots,\vec{x}(t_k),\vec{y}_{1:k})$
of the same dimensionality as $\set{T}_0$
\begin{equation}
  p(\vectheta\,|\,\vec{x}(t_1),\ldots,\vec{x}(t_k),\vec{y}_{1:k})
  = p(\vectheta\,|\,\set{T}_k),
\end{equation}
such that there exists an algorithm $\Phi(\cdot)$ that can be used for
efficiently performing the update
\begin{equation}
  \set{T}_k = \Phi\left( \set{T}_{k-1}, \vec{x}(t_k), \vec{y}_{k} \right).
\end{equation}
Further assume that the marginal likelihood
\begin{equation}
  p(\vec{y}_k\,|\,\vec{x}(t_k), \set{T}_{k-1})
  = \int p(\vec{y}_k\,|\,\vec{x}(t_k),\vectheta) \, p(\vectheta\,|\,\set{T}_{k-1}) \,
  \diff \vectheta,
\end{equation}
can be efficiently evaluated. The above conditions are met, for
example, when for fixed $\vec{x}(t_k)$ the distribution
$p(\vectheta\,|\,\set{T}_{k-1})$ is conjugate for the likelihood
$p(\vec{y}_k\,|\,\vec{x}(t_k),\vectheta)$ with respect to $\vectheta$.

The resulting algorithm is now the following:
\begin{algorithm}[CDRB-SIR II] \label{alg:cdrbsir2} Given the
  importance process, a weighted set of samples $\{
  \vec{x}^{(i)}_{k-1}, \set{T}^{(i)}_{k-1}, w^{(i)}_{k-1} \}$ and the
  new measurement $\vec{y}_k$, a single step of {\em
    continuous-discrete Rao-Blackwellized SIR with static parameters}
  can be performed as follows:
\begin{enumerate}
\item Simulate the importance process, scaled importance process
  and log-likelihood ratio as in Algorithm \ref{alg:cdsir1} or
  \ref{alg:cdsir2}. This results in the sample set $\{
  \vec{x}^{(i)}_k \}$ and likelihood ratios $\{ Z^{(i)}_k \}$.

\item For each $i$ compute new sufficient statistics
\begin{equation}
  \set{T}^{(i)}_k = \Phi\left( \set{T}^{(i)}_{k-1}, \vec{x}^{(i)}_k, \vec{y}_{k} \right),
\end{equation}
evaluate the importance weights as
\begin{equation}
\begin{split}
  w^{(i)}_{k} &= w^{(i)}_{k-1} \, Z^{(i)}_k \,
  p(\vec{y}_k\,|\,\vec{x}^{(i)}_k, \set{T}^{(i)}_{k-1}),
\end{split}
\end{equation}
and re-normalize the weights to sum to unity.

\item If the effective number of particles is too low, perform
  resampling.
\end{enumerate}
\end{algorithm}
Actually, the sufficient statistics could be functionals of the whole
state trajectory, in which case they could be simulated together with
the state.

\section{Numerical Simulations}
In this section the continuous-discrete sequential importance sampling
is applied to estimation of partially measured simple pendulum which
is distorted by a random noise term and to estimation of the spread of
an infectious disease. Several other applications and the more details
on the presented applications can be found in the doctoral
dissertation of \cite{Sarkka:2006a}.

\subsection{Simple Pendulum with Noise}
\label{sec:simu}
The stochastic differential equation for the angular position of a
simple pendulum \citep{Alonso+Finn:1980}, which is distorted by random
white noise accelerations $w(t)$ with spectral density $q$ can be
written as
\begin{equation}
    \frac{\diff^2 x}{\diff t^2} + a^2 \, \sin(x) = w(t).
\end{equation}
where $a$ is the angular velocity of the (linearized) pendulum. If we
define the state as $\vec{x} = (x~\diff x / \diff t)^T$ and change to
state space form and to the integral equation notation in terms of
Brownian motion, the model can be written as
\begin{equation}
  \begin{split}
    \frac{\diff x_1}{\diff t} &= x_2 \\
    \diff x_2 &= -a^2 \, \sin(x_1) \, \diff t
    + \diff \beta,
  \end{split}
\label{eq:pend_sde}
\end{equation}
where $\beta(t)$ has the diffusion coefficient $q$, which is model of
the form \eqref{eq:lh2model}. Assume that the state of the pendulum is
measured once per unit time and the measurements are corrupted by
Gaussian measurement noise with an unknown variance $\sigma^2$. A
suitable model in this case is
\begin{equation}
  \begin{split}
        y_k &\sim \N(x_1(t_k),\sigma^2) \\
   \sigma^2 &\sim \text{Inv-$\chi^2$}(\nu_0,\sigma_0^2),
  \end{split}
  \label{eq:pend_meas}
\end{equation}
This is now a model with an unknown static parameter as discussed in
Section \ref{sec:cdrbsir2}.

The importance process can be now formed by the continuous-discrete
extended Kalman filter (EKF) \citep[see,
e.g., ][]{Jazwinski:1970,Gelb:1974} and the result is a 2-dimensional
Gaussian approximation for the joint distribution of the state
$\vec{x}(t_k) = (x_1(t_k)~x_2(t_k))^T$. Forming this approximation
requires that the variance $\sigma^2$ is assumed to be known, but
fortunately a very rough approximation based on the estimated
$\sigma_k^2$ is enough in practice. In that case the EKF based
approximation can be constructed as follows:
\begin{enumerate}
\item Assume that the posterior distribution of a particle
  $\vec{x}^{(i)}(t)$ is approximately Gaussian
\begin{equation}
  \vec{x}^{(i)}(t)\,|\,\vec{y}_{1:k-1} \sim \N(\vec{m}(t),\mat{P}(t)).
\end{equation}
Note that immediately after a measurement, a single sampled particle
actually has a Dirac delta distribution, which also is a (degenerate)
Gaussian distribution.
\item By forming a first order Taylor series expansion to the right
  hand side of the equation \eqref{eq:pend_sde} we get that after a
  sufficiently small time interval $\delta t$ the state mean and
  covariance can be approximated as
\begin{equation}
  \begin{split}
    \vec{m}(t + \delta t)
    &= \vec{m}(t) + \vec{f}(\vec{m}(t)) \, \delta t \\
    \mat{P}(t + \delta t)
    &= \mat{P}(t) + \left[ \mat{F}(\vec{m}(t)) \, \mat{P}(t) + \mat{P}(t) \,
    \mat{F}^T(\vec{m}(t)) + \mat{Q} \right] \, \delta t, \\
  \end{split}
\end{equation} 
where $\vec{f}(\vec{x}) = (x_2~-a^2 \sin(x_1))^T$, $\mat{F}(\vec{x})$
is the Jacobian matrix of $\vec{f}(\vec{x})$ and $\mat{Q} =
\diag(0~q)$.

\item We may now form Gaussian approximation to the state at time $t +
  \delta t$ with the mean and covariance above. If we continue this
  process recursively and take limit $\delta t \rightarrow 0$, we get
  that we may approximate the process as Gaussian process with mean
  and covariance
\begin{equation}
  \begin{split}
    \frac{\diff \vec{m}(t)}{\diff t}
    &= \vec{f}(\vec{m}(t)) \\
    \frac{\diff \mat{P}(t)}{\diff t}
    &= \mat{F}(\vec{m}(t)) \, \mat{P}(t) + \mat{P}(t) \,
    \mat{F}^T(\vec{m}(t)) + \mat{Q}, \\
  \end{split}
\label{eq:pend_ekf}
\end{equation} 
\end{enumerate}
The above result states that between the measurements we can
approximate the mean and covariance of the process \eqref{eq:pend_sde}
by integrating the deterministic differential equations
\eqref{eq:pend_ekf}. The result is a Gaussian process, that is, a
Gaussian approximation to the state process at any instance of time.

The importance process can be now constructed as follows. For each
particle $i$ do the following:
\begin{enumerate}
\item Solve the approximate predicted mean and covariance at time
  $t_k$ from the differential equations \eqref{eq:pend_ekf} by
  starting from initial conditions $\vec{m}(t_{k-1}) =
  \vec{x}^{(i)}_{k-1}$, $\mat{P}(t_{k-1}) = \mat{0}$.

\item Assuming that $\sigma^2$ is known, the approximate joint
  distribution of the state and measurement is Gaussian and thus we
  can compute the posterior distribution of the state in closed form.
\end{enumerate}
If the resulting approximate marginal posterior mean and covariance of
$x_2(t_k)$ are $m_{2,k}$ and $P_{22,k}$, then a suitable importance
process is (assuming that sampling interval is $\Delta t$)
\begin{equation}
  \begin{split}
    \frac{\diff s_1}{\diff t} &= s_2 \\
    \diff s_2 &= \left( \frac{m_{2,k} - x_{2,k-1}}{\Delta t} \right)
    \, \diff t + \sqrt{\frac{P_{22,k}}{q \, \Delta t}} \, \diff \beta,
  \end{split}
  \label{eq:pend_s}
\end{equation}
with initial conditions
\begin{equation}
  \begin{split}
     s_1(t_{k-1}) &= x_{1,k-1} \\
     s_2(t_{k-1}) &= x_{2,k-1}.
  \end{split}
\end{equation}
The equations for the scaled importance process can be now written as
\begin{equation}
  \begin{split}
    \frac{\diff s^*_1}{\diff t} &= s^*_2 \\
    \diff s^*_2 &=
     \left( \sqrt{\frac{q}{P_{22,k} \, \Delta t}} \right)
     \left( m_{2,k} - x_{2,k-1} \right)
    \, \diff t + \diff \beta,
  \end{split}
  \label{eq:pend_sp}
\end{equation}
with initial conditions
\begin{equation}
  \begin{split}
    s^*_1(t_{k-1}) &= x_{1,k-1} \\
    s^*_2(t_{k-1}) &= x_{2,k-1},
  \end{split}
\end{equation}
The full state of the algorithm at time step $k-1$ consists of the set
of particles
\begin{equation}
  \{
    w^{(i)}_{k-1},
    x^{(i)}_{1,k-1},
    x^{(i)}_{2,k-1},
    \nu^{(i)}_{k-1},
    \sigma^{2,(i)}_{k-1}
  \}
\end{equation}
where $w^{(i)}_{k-1}$ is the importance weight, $x^{(i)}_{1,k-1},
x^{(i)}_{2,k-1}$ is the state of the pendulum, and $\nu^{(i)}_{k-1},
\sigma^{2,(i)}_{k-1}$ are the sufficient statistics of the
variance parameter.

\begin{figure}
\centering
\includegraphics{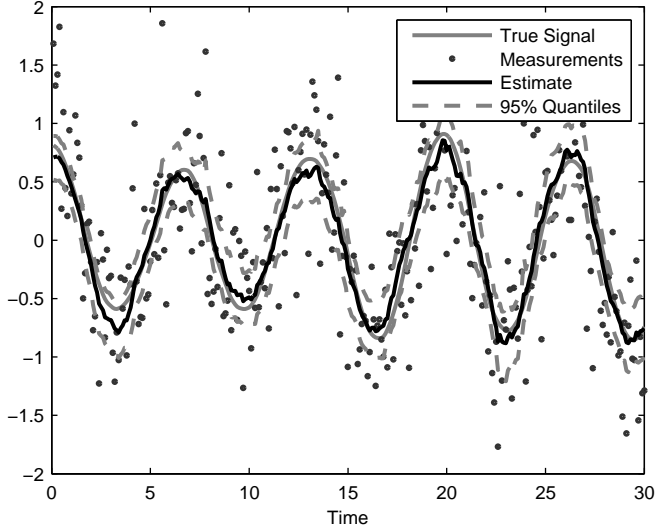}
\caption{The result of applying con\-tin\-u\-ous-discrete particle
filter with EKF proposal to a simulated noisy pendulum data.}
\label{fig:pend_res}
\end{figure}

Figure \ref{fig:pend_res} shows the result of applying the
con\-tin\-u\-ous-discrete particle filter with EKF proposal and $1000$
particles to a simulated data. The data was generated from the noisy
pendulum model with process noise spectral density $q = 0.01$, angular
velocity $a=1$ and the sampling step size was $\Delta t = 0.1$. The
estimate can be seen to be quite close to the true signal.


In the simulation, the true measurement variance was $\sigma^2 =
0.25$. The prior distribution used for the unknown variance parameter
was $\sigma^2 \sim \text{Inv-}\chi^2(2,0.2)$.

\begin{figure}
\centering
\includegraphics{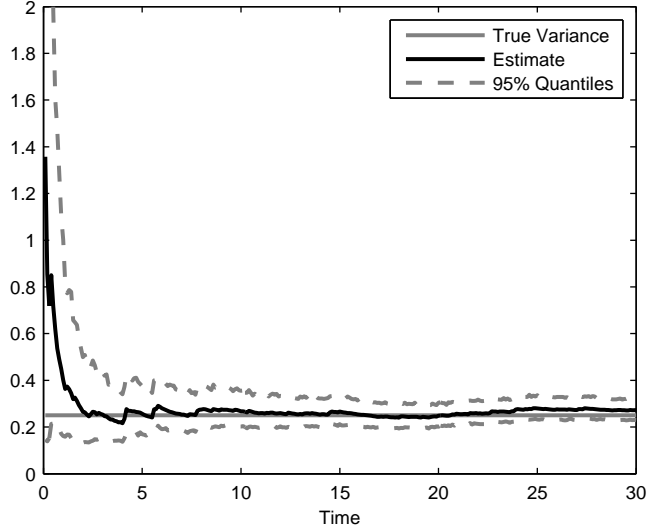}
\caption{The evolution of variance distribution in the noisy pendulum problem.}
\label{fig:pend_var}
\end{figure}

The evolution of the posterior distribution of the variance parameter
is shown in the Figure \ref{fig:pend_var}. In the beginning the
uncertainty about the variance is higher, but the distribution quickly
concentrates to the neighborhood of the true value.

%
\subsection{Spread of Infectious Diseases} \label{sec:disease}
%

The classic model for the dynamics of infectious diseases is the SIR
model (The model is called the SIR model, because the variables
$X(t)$, $Y(t)$, and $Z(t)$ denote the susceptible, infective and
removed compartments and for this reason are often denoted as $S(t)$,
$I(t)$, and $R(t)$, respectively)
\citep{Kermack+McKendrick:1927,Anderson+May:1991,Murray:1993,Hethcote:2000},
which is valid for sufficiently large $N$:
\begin{align}
  \diff X / \diff t
  &= -b \, Y \, X / N,      \qquad &X(0) &= X_0, \\
  \diff Y / \diff t
  &= b \, Y \, X / N - g \, Y,  \qquad &Y(0) &= Y_0, \\
  \diff Z / \diff t
  &= g \, Y,               \qquad &Z(0) &= Z_0,
\end{align}
where $X(t)$ is the number of susceptibles at time $t$, $X_0 \ge 0$ is
the initial number of susceptibles, $Y(t)$ is the number of infectives
who are capable of transmitting the infection, $Y_0 \ge 0$ is the
initial number of infectives, $Z(t)$ is the number of recovered or
dead individuals which cannot be infected anymore, $Z_0 \ge 0$ is the
initial number of individuals in this class, $N = X(t) + Y(t) + Z(t)$
is the (constant) total number of individuals, $b$ is the contact rate
which determines the rate of individuals moving from susceptible class
to infectious class, and $g$ is the waiting time parameter such that
$1/g$ is the average length of the infectious period.

If we model the contact number $\sigma = b/g$ as the exponential of
the Brownian motion, then the stochastic equations for the proportions
of individuals in each class can be written as \citep{Sarkka:2006a}:
\begin{equation}
\begin{split}
  \diff x / \diff t
  &= -g \, \exp(\lambda) \, y \, x \\
  \diff y / \diff t
  &= g \, \exp(\lambda) \, y \, x - g \, y \\
  \diff \lambda
  &= q^{1/2} \, \diff \beta,
\end{split}
\label{eq:ssir}
\end{equation}
where $\beta(t)$ is a standard Brownian motion and $\lambda = \ln
\sigma$.

A suitable initial distribution for $x(0)$ and $y(0)$ is
\begin{align}
  y(0) &\sim \mathrm{Beta}(\alpha_y,\beta_y), \\
  x(0) &= 1 - y(0),
\end{align}
where $\beta_y \gg \alpha_y$. The initial conditions $z(0)$ can be
assumed to be zero without loss of generality.

In the classical SIR model the values $X(t)$, $Y(t)$ and $Z(t)$ are
not restricted to integer values, and thus they cannot be interpreted
as counts as such. A sensible stochastic interpretation of these
values is that they are the average numbers of individuals in each
class and the actual numbers of individuals are Poisson distributed
with these means. 

Assume that the number of dead individuals is recorded. Then the
number of the dead individuals $d_k$ on time period $[t_{k-1},t_k]$
has the distribution
  \begin{equation}
     p(d_k\,|\,\{ x(\tau),y(\tau) : 0 \le \tau \le t_k \},N)
     = \mathrm{Poisson}(d_k\,|\,N \, \theta_k),
  \label{eq:sir_d}
  \end{equation}
  where
  \begin{equation}
    \theta_k = x(t_{k-1})-x(t_{k})+y(t_{k-1})-y(t_{k}).
  \end{equation}
The population size $N$ is unknown and it can be modeled as having a
Gamma prior distribution
\begin{equation}
  p(N) = \mathrm{Gamma}(N\,|\,\alpha_0,\beta_0),
\end{equation}
with some suitably chosen $\alpha_0$ and $\beta_0$. As shown in
\citep{Sarkka:2006a} this model is now of such form that it is
possible to integrate out the population size $N$ from the equations
and the Algorithm \ref{alg:cdrbsir2} can be applied.

The continuous-discrete SIR filter was applied to the classical Bombay
plague data presented in \citep{Kermack+McKendrick:1927}. An EKF based
Gaussian process approximation was used as the importance process
\citep[see, ][for details]{Sarkka:2006a} and $10000$ particles was
used. The prior distribution for proportion of initial infectives was
$\text{Beta}(1,100)$. The population size prior was
$\text{Gamma}(10,0.001)$. The waiting time parameter was assumed to be
$g=1$. The prior distribution for $\lambda(0)$ was $\N(\ln(5),4)$. The
diffusion coefficient of the Brownian motion was $q=0.001$.

\begin{figure}[htb!]
\begin{center}
\includegraphics{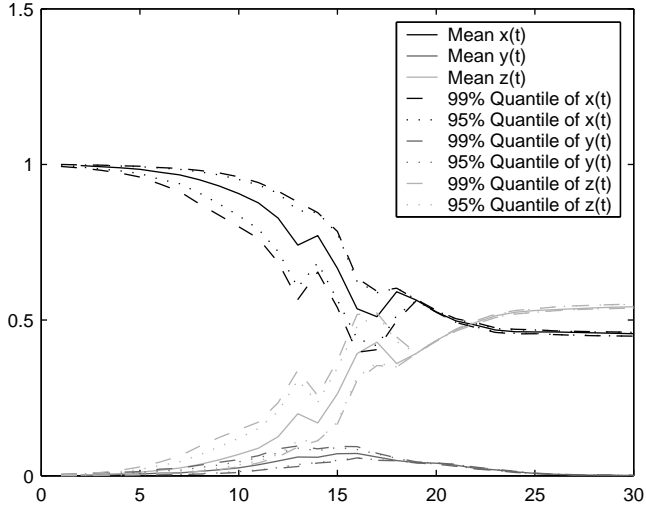}
\end{center}
\caption{Filtered estimates of values of $x(t)$, $y(t)$, and $z(t)$ from
  the Bombay data.}
\label{fig:bombay_xy}
\end{figure}

The final filtered estimates of the histories of $x(t)$, $y(t)$, and
$z(t)$ are shown in Figure \ref{fig:bombay_xy}. These estimates are
filtered estimates, that is, they are conditional to the previously
observed measurements only. That is, the estimate on week $t$ is the
estimate that could be actually computed on week $t$ without any
knowledge of the future observations. The estimates look quite much as
what would be expected: the proportion of susceptibles $x(t)$
decreases in time and the number of infectives $y(t)$ increases up to
a maximum and then decreases to zero. However, these estimated values
are not very useful themselves. The reason for this is that, for
example, the value $x_{\infty}$ which is the remaining value of
susceptibles in the end depends on the choice of $g$ and other prior
parameters. That is, these estimated values are not absolute in the
sense that their values depend heavily on the prior assumptions.

\begin{figure}[htb!]
\begin{center}
\includegraphics{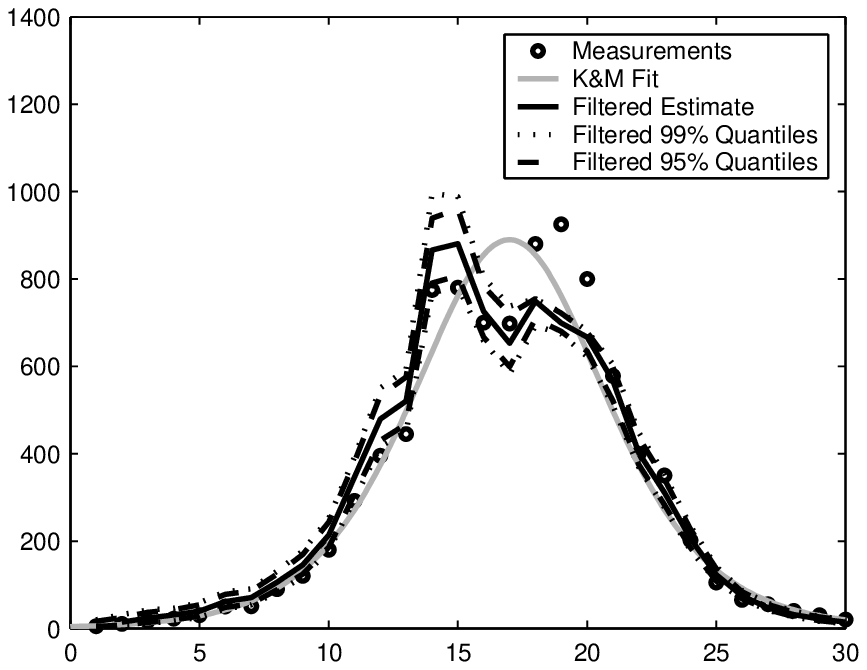}
\end{center}
\caption{Filtered estimate of $\diff Z / \diff t$ from the Bombay data.
  The estimate of \citep{Kermack+McKendrick:1927} is also shown for
  comparison.}
\label{fig:bombay_drdtf}
\end{figure}

Much more informative quantity is the value $\diff Z / \diff t$, whose
filtered estimate is shown in Figure \ref{fig:bombay_drdtf}.  The
classical estimate presented in \citep{Kermack+McKendrick:1927} is
also shown. The SIR filter estimate can be seen to differ a bit from
the classical estimate, but still both the estimates look quite much
like what would be expected. Note that the classical estimate is based
on all measurements, whereas the filtered estimate is based on
observations made up to that time only. That is, the filter estimate
could be actually computed on week $t$, but the classical estimate
could not.

\begin{figure}[htb!]\begin{center}
\includegraphics{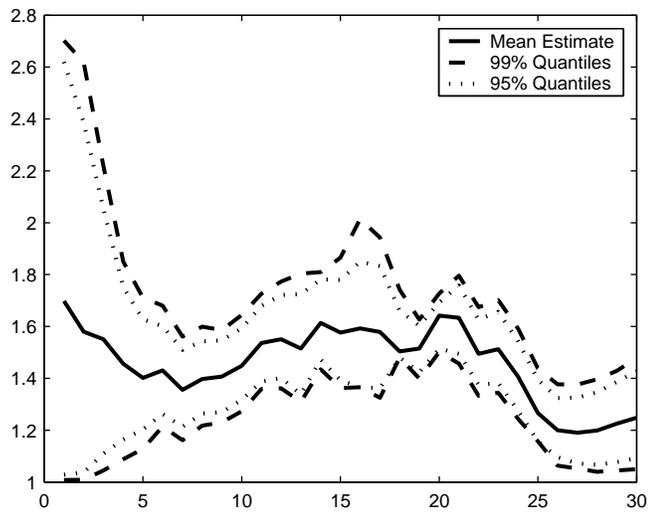}
\end{center}
\caption{Bombay plague: Filtered estimate of value $\sigma(t)$.}
\label{fig:bombay_sigma}
\end{figure}

The filtered estimates of values $\sigma(t)$ are shown in Figure
\ref{fig:bombay_sigma}. The value can be seen to vary a bit on time,
but the estimated expected value remains on the range $[1.4,1.8]$ all
the time. As can be seen from the figure, according to the data the
value of $\sigma(t)$ is not constant. This is not surprising, because
the spatial and other unknown effects are not accounted at all in the
classical SIR model and these effects typically affect the number of
contacts.

\begin{figure}[htb!]
\begin{center}
\includegraphics{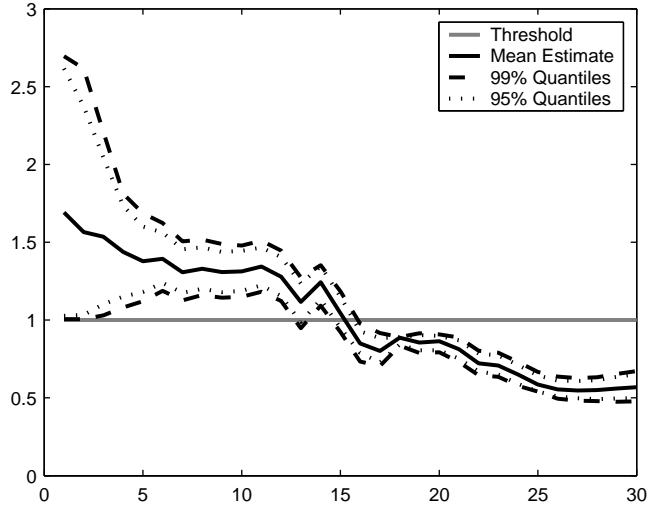}
\end{center}
\caption{Bombay plague: Filtered estimate of value $x(t) \, \sigma(t)$.}
\label{fig:bombay_ssi}
\end{figure}

A very useful indicator value is $\sigma(t) \, x(t)$, whose filtered
estimate is shown in Figure \ref{fig:bombay_ssi}. In the deterministic
SIR model with constant $\sigma$ this indicator defines the asymptotic
behavior of the epidemic \citep[see, e.g.,][]{Hethcote:2000}: If
$\sigma x(t) \le 1$ then the number of infectives will decrease to
zero as $t \rightarrow \infty$. If $\sigma x(t) > 1$ then the number
of infectives will first increase up to a maximum and then decrease to
zero. As can be seen from the Figure \ref{fig:bombay_ssi} the filtered
estimate of the indicator value goes below 1 just after the maximum
somewhere between weeks 15--16, which can be seen in Figure
\ref{fig:bombay_drdtf}. That is, the estimated value of $\sigma(t) \,
x(t)$ could be used as an indicator, which tells if the epidemic is
over or not.

\begin{figure}[htb!]
\begin{center}
\includegraphics{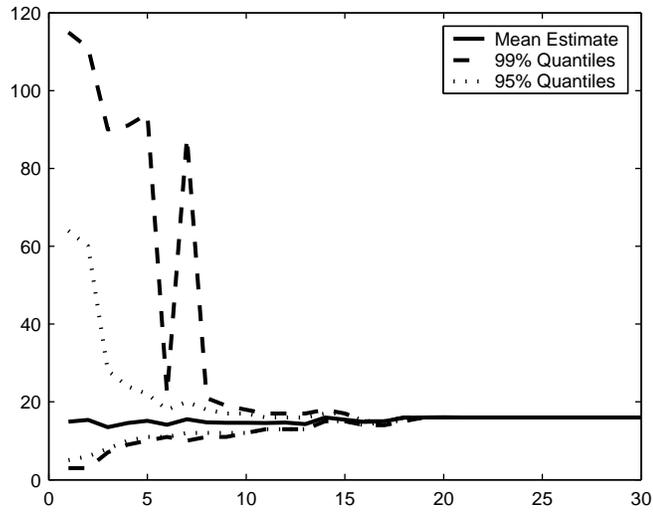}
\end{center}
\caption{Bombay plague: Filtered estimate of time of maximum of epidemic.}
\label{fig:bombay_max}
\end{figure}

Using the particles it is also possible to predict ahead to the future
and estimate the time when the maximum of the epidemic will be
reached. The estimate computed from the filtering result is shown in
the Figure \ref{fig:bombay_max}. Again, the estimates are filtered
estimates and the estimate on week $t$ could be actually computed on
week $t$, because it depends only on the counts observed up to that
time. The filtered estimate can be seen to quickly converge to the
values near the correct maximum on weeks 15--16. It is interesting to
see that the prediction is quite accurate already around the week 10,
which is far before reaching the actual maximum. If this kind of
prediction had been done on, for example, week 10 of the disease, it
would have predicted the time of actual epidemic maximum quite
accurately. After the maximum has been observed, the estimate quickly
converges to a constant value, which according to the Figure
\ref{fig:bombay_drdtf} is likely to be near the true maximum.

\begin{figure}[htb!]
\begin{center}
\includegraphics{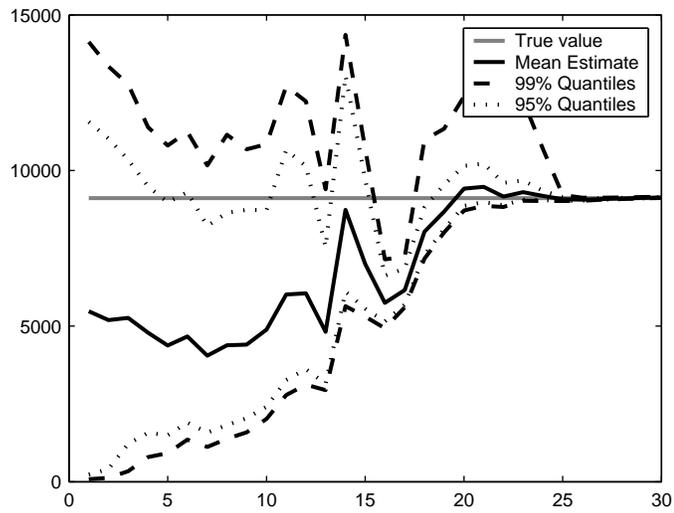}
\end{center}
\caption{Bombay plague: Filtered estimated of number of deaths.}
\label{fig:bombay_nd}
\end{figure}

A very useful estimate is also the expected total number of deaths
caused by the epidemic. This can be computed from the filtered
estimates and the result is shown in Figure \ref{fig:bombay_nd}. In
the beginning the estimate is very diffuse, but after maximum has been
reached the estimate converges near the correct value. The estimate is
a bit less than the observed value long before reaching the maximum,
which might be due to existence of two maximums in the observed data
(see, Figure \ref{fig:bombay_drdtf}).  Because the second maximum is
not predicted by the model, the extra number of deaths caused by it
cannot be seen in the predictions.

\section{Discussion}
\label{sec:discussion}

The importance processes used in the continuous-discrete particle
filtering examples are very simple and better alternatives definitely
exists. In principle, the optimal importance process in the
con\-tin\-u\-ous-discrete particle filtering case would have the same
law as the smoothing solution. Thus, constructing the importance
process based on the smoothing solution instead of linearly
interpolated filtering solutions, as in this article, could lead to
more efficient particle filtering methods. In some cases it could be
possible to construct a process, which would have exactly the same law
as the optimal importance process.

A weakness in the con\-tin\-u\-ous-discrete particle filtering
framework is that the importance process has to be scaled before
sampling. In practice, this restricts the possible forms of importance
processes to those having the same dispersion matrix as the original
process. However, this a bit more general case with explicit scaling
is treated here instead of requiring $\mat{L}(t) = \mat{B}(t)$,
because this leaves more room to the possibility that maybe the
equations could be modified such that the scaling of the importance
process would not be needed.

Another weakness of the framework is that the time-discretization
introduces biasedness to the estimation. The time-discretization is
due to the usage of numerical integration methods for SDEs, which use
discretization in time. However, there exists method for simulating
SDEs without time-discretization \citep{Beskos:2006} and maybe by
using this kind of methods this biasedness could be eliminated.

The con\-tin\-u\-ous-discrete sequential importance resampling
framework could be extended to the case of stochastic differential
equations driven by more general martingales, for example, general
L\'evy processes such as compound Poisson processes
\citep{Applebaum:2004}. This would allow modeling of sudden changes in
signals. This extension could be possible by simply replacing the
Brownian motion in the Girsanov theorem by a more general martingale.

It could be possible to generalize the con\-tin\-u\-ous-discrete
sequential importance sampling framework to con\-tin\-u\-ous-time
filtering problems. Then the extended Kalman-Bucy filter or the
unscented Kalman-Bucy filter \citep{Sarkka:2006a,Sarkka:2007c} could
be used for forming the importance process and the actual filtering
result would be formed by weighting the importance process samples
properly.

The likelihood ratio expressions in Theorems \ref{the:trans1} and
\ref{the:trans2} have an interesting connection to the variational
method considered in article
\citep{Archambeau+Cornford+Opper+Shawe-Taylor:2007}. If we select the
processes as
\begin{align}
  \diff \vec{x}(t) &= \vec{f}(\vec{x}(t),t) \, \diff t
                 + \sqrt{\Sigma} \, \diff \vecbeta(t) \\
  \diff \vec{s}(t) &= \vec{f}_L(\vec{s}(t),t) \, \diff t
                 + \sqrt{\Sigma} \, \diff \vecbeta(t),
\end{align}
where $\vec{x}(t)$ is a process with density $p(\cdot)$ and $\vec{s}(t)$ is
a process with density $q(\cdot)$, then by taking the expectation of
negative logarithm of \eqref{eq:lh1z} we get the expression for the
KL-divergence between $q$ and $p$:
\begin{equation}
\begin{split}
  \mathrm{KL}[q\,|\,p]
  &= \E\Big[ \frac{1}{2}
    \int_0^t \big\{\vec{f}(\vec{s}(t),t) - \vec{f}_L(\vec{s}(t),t)\big\}^T \Sigma^{-1} 
   \big\{\vec{f}(\vec{s}(t),t) - \vec{f}_L(\vec{s}(t),t)\big\} \,
    \diff t \Big],
\end{split}
\end{equation}
which is exactly the expression obtained heuristically in
\citet{Archambeau+Cornford+Opper+Shawe-Taylor:2007}. Thus the
extensions to singular models would also apply to that method.

\section{Conclusions}
\label{sec:conclusion}

In this article, a new class of methods for con\-tin\-u\-ous-discrete
sequential importance sampling (particle filtering) has been
presented. These methods are based on transformations of probability
measures using the Girsanov theorem. The new methods are applicable to
a general class of models. In particular, they can be applied to many
models with singular dispersion matrices, unlike many previously
proposed measure transformation based sampling methods. The new
methods have been illustrated in a simulated problem, where both the
implementation details of the algorithms and the simulation results
have been reported. The methods have also been applied to estimation
of the spread of an infectious disease based on counts of dead
individuals.

The classical con\-tin\-u\-ous-discrete extended Kalman filter as well
as the recently developed con\-tin\-u\-ous-discrete unscented Kalman
filter can be used for forming importance processes for the new
con\-tin\-u\-ous-discrete particle filters. This way the efficiency of
the Gaussian approximation based filters can be combined with the
accuracy of the particle approximations. Closed form marginalization
or Rao-Blackwellization can be applied if the model is conditionally
Gaussian or if the model contains unknown static parameters and has a
suitable conjugate form. In most cases Rao-Blackwellization leads to a
significant improvement in the efficiency of the particle filtering
algorithm.

\appendix

\section{Likelihood Ratios of SDEs}
\label{app:lh}
%

In the computation of the likelihood ratios of stochastic differential
equations we need a slightly generalized version of the Girsanov
theorem \citep{Kallianpur:1980,Karatzas+Shreve:1991,Oksendal:2003}.
The generalized theorem can be obtained, for example, as a special
case from the theorems presented in \citet{Delyon+Hu:2006}.

\begin{theorem}[Girsanov] \label{the:girsanov}
Let $\vecbeta=(\beta_1,\ldots,\beta_d)$ be a Brownian
motion with diffusion matrix $\mat{Q}(t)$ under the probability
measure $\mea{P}$. Let $\vectheta : \Omega\times\spc{R}_+ \mapsto
\spc{R}^{d}$ be an adapted process such that the process $Z$ defined
as
\begin{equation}
Z(t) = \exp\left\{
\int_0^t \vectheta^T(t) \diff \vecbeta(t) - \frac{1}{2}
\int_0^t \vectheta^T(t) \, \mat{Q}(t) \, \vectheta(t) \diff t \right\},
\end{equation}
satisfies $\E[Z(t)]=1$. Then the process
\begin{equation}
\diff \tilde{\vecbeta}(t)
= \diff \vecbeta(t) - \mat{Q}(t) \, \vectheta(t) \diff t
\end{equation}
is a Brownian motion with diffusion matrix $\mat{Q}(t)$ under the
probability measure $\tilde{\mea{P}}$ defined via the relation
\begin{equation}
\E\left[\frac{\diff \tilde{\mea{P}}}{\diff\mea{P}}\,\bigg|\,\alg{F}_t\right] 
= Z(t),
\end{equation}
where $\alg{F}_t$ is the natural filtration of the Brownian motion
$\vecbeta(t)$.
\end{theorem}

\begin{proof}
See, for example, \citet{Delyon+Hu:2006}.
\end{proof}

\begin{theorem}[Transformation of SDE Solutions I] \label{the:trans1}
Let
\begin{align}
  \diff \vec{x}(t) &= \vec{f}(\vec{x}(t),t) \, \diff t
                 + \mat{L}(t) \, \diff \vecbeta(t),
  \qquad \vec{x}(0) = x_0 \label{eq:lhsde1} \\
  \diff \vec{s}(t) &= \vec{g}(\vec{s}(t),t) \, \diff t
                 + \mat{B}(t) \, \diff \vecbeta(t),
  \qquad \vec{s}(0) = x_0, \label{eq:lhsde2}
\end{align}
where $\vecbeta(t)$ is a Brownian motion with diffusion matrix
$\mat{Q}(t)$ with respect to measure $\mea{P}$. Let $\alg{F}_t$ be its
natural filtration. The matrices $\mat{L}(t)$ and $\mat{B}(t)$ are
assumed to be invertible for all $t$. Now the process $\vec{s}^*(t)$
defined as
\begin{equation}
  \diff \vec{s}^* = \mat{L}(t) \, \mat{B}^{-1}(t) \, \diff \vec{s},
  \qquad \vec{s}(0) = \vec{x}_0
\label{eq:sstar}
\end{equation}
is a weak solution to the Equation \eqref{eq:lhsde1} under the measure
$\tilde{\mea{P}}$ defined by the relation
\begin{equation}
  \E\left[\frac{\diff \tilde{\mea{P}}}
  {\diff\mea{P}}\,\bigg|\,\alg{F}_t\right] 
  = Z(t).
\label{eq:lh1lhr}
\end{equation}
where
\begin{equation}
\begin{split}
  Z(t) &=
  \exp\Big[
    \int_0^t \big\{\vec{f}(\vec{s}^*(t),t)
  - \mat{L}(t) \, \mat{B}^{-1}(t) \,
    \vec{g}(\vec{s}(t),t)\big\}^T \, \mat{L}^{-T}(t) \,
    \mat{Q}^{-1}(t) \, \diff \vecbeta(t)
  \\
  &\qquad - \frac{1}{2}
    \int_0^t \big\{\vec{f}(\vec{s}^*(t),t)
  - \mat{L}(t) \, \mat{B}^{-1}(t) \, \vec{g}(\vec{s}(t),t)\big\}^T \\
  &\quad \qquad \times \big\{ \mat{L}(t) \, \mat{Q}(t) \, \mat{L}^T(t) \big\}^{-1} 
   \big\{\vec{f}(\vec{s}^*(t),t)
  - \mat{L}(t) \, \mat{B}^{-1}(t) \, \vec{g}(\vec{s}(t),t)\big\} \,
    \diff t
  \Big]
\end{split}
\label{eq:lh1z}
\end{equation}
\end{theorem}

\begin{proof}
By substituting the expression \eqref{eq:lhsde2} into Equation
\eqref{eq:sstar}, solving for $\diff \vecbeta(t)$, we get
\begin{equation}
\diff \vecbeta(t) = 
  \mat{L}^{-1}(t) \, \diff \vec{s}^* -
  \mat{B}^{-1}(t) \, \vec{g}(\vec{s}(t),t) \, \diff t.
\end{equation}
If we now define
\begin{equation}
  \vectheta(t) = 
    \mat{Q}^{-1}(t) \, \mat{L}^{-1}(t) \, \vec{f}(\vec{s}^*(t),t)
  - \mat{Q}^{-1}(t) \, \mat{B}^{-1}(t) \, \vec{g}(\vec{s}(t),t),
\end{equation}
then under the measure $\tilde{\mea{P}}$ defined by \eqref{eq:lh1lhr}
and \eqref{eq:lh1z} with the process $\vectheta(t)$ defined as above,
the following process is a Brownian motion with diffusion matrix
$\mat{Q}(t)$:
\begin{equation}
\begin{split}
\diff \tilde{\vecbeta}(t)
&= \diff \vecbeta(t) - \mat{Q}(t) \, \vectheta(t) \diff t \\
&= \mat{L}^{-1}(t) \, \diff \vec{s}^* -
   \mat{B}^{-1}(t) \, \vec{g}(\vec{s}(t),t) \, \diff t \\
&- \mat{Q}(t) \, \mat{Q}^{-1}(t) \,
   \mat{L}^{-1}(t) \, \vec{f}(\vec{s}^*(t),t) 
    \, \diff t
 + \mat{Q}(t) \, \mat{Q}^{-1}(t) \,
   \mat{B}^{-1}(t) \, \vec{g}(\vec{s}(t),t)
    \, \diff t\\
&= \mat{L}^{-1}(t) \, \diff \vec{s}^* 
 - \mat{L}^{-1}(t) \, \vec{f}(\vec{s}^*(t),t) \, \diff t
\end{split}
\end{equation}
By rearranging we get that
\begin{equation}
\diff \vec{s}^* = 
\vec{f}(\vec{s}^*(t),t) \, \diff t + \mat{L}(t) \, \diff \tilde{\vecbeta}(t)
\end{equation}
and thus the result follows. The explicit expression for the
likelihood ratio is given as follows:
\begin{equation}
\begin{split}
  Z(t) &=
  \exp\Big[
    \int_0^t \big\{\mat{Q}^{-1}(t) \,
    \mat{L}^{-1}(t) \, \vec{f}(\vec{s}^*(t),t)
  - \mat{Q}^{-1}(t) \, \mat{B}^{-1}(t) \,
    \vec{g}(\vec{s}(t),t)\big\}^T \, \diff \vecbeta(t)
  \\
  &- \frac{1}{2}
    \int_0^t \big\{\mat{Q}^{-1}(t) \,
    \mat{L}^{-1}(t) \, \vec{f}(\vec{s}^*(t),t)
  - \mat{Q}^{-1}(t) \, \mat{B}^{-1}(t) \, \vec{g}(\vec{s}(t),t)\big\}^T \\
  &\times \mat{Q}(t) \big\{\mat{Q}^{-1}(t) \,
    \mat{L}^{-1}(t) \, \vec{f}(\vec{s}^*(t),t)
  - \mat{Q}^{-1}(t) \, \mat{B}^{-1}(t) \, \vec{g}(\vec{s}(t),t)\big\} \,
    \diff t
  \Big] \\
  &=
  \exp\Big[
    \int_0^t \big\{\vec{f}(\vec{s}^*(t),t)
  - \mat{L}(t) \, \mat{B}^{-1}(t) \,
    \vec{g}(\vec{s}(t),t)\big\}^T \, \mat{L}^{-T}(t) \,
    \mat{Q}^{-1}(t) \, \diff \vecbeta(t)
  \\
  &\qquad - \frac{1}{2}
    \int_0^t \big\{\mat{L}^{-1}(t) \, \vec{f}(\vec{s}^*(t),t)
  - \mat{L}(t) \, \mat{B}^{-1}(t) \, \vec{g}(\vec{s}(t),t)\big\}^T \\
  &\quad \qquad \times \big\{ \mat{L}^{-T}(t) \, \mat{Q}^{-1}(t) \,
    \mat{Q}(t) \mat{Q}(t)^{-1} \, \mat{L}^{-1}(t) \big\} \\
  &\quad \qquad \times \big\{\vec{f}(\vec{s}^*(t),t)
  - \mat{L}(t) \, \mat{B}^{-1}(t) \, \vec{g}(\vec{s}(t),t)\big\} \,
    \diff t
  \Big] \\
  &=
  \exp\Big[
    \int_0^t \big\{\vec{f}(\vec{s}^*(t),t)
  - \mat{L}(t) \, \mat{B}^{-1}(t) \,
    \vec{g}(\vec{s}(t),t)\big\}^T \, \mat{L}^{-T}(t) \,
    \mat{Q}^{-1}(t) \, \diff \vecbeta(t)
  \\
  &\qquad - \frac{1}{2}
    \int_0^t \big\{\vec{f}(\vec{s}^*(t),t)
  - \mat{L}(t) \, \mat{B}^{-1}(t) \, \vec{g}(\vec{s}(t),t)\big\}^T \\
  &\quad \qquad \times \big\{ \mat{L}(t) \, \mat{Q}(t) \, \mat{L}^T(t) \big\}^{-1} 
   \big\{\vec{f}(\vec{s}^*(t),t)
  - \mat{L}(t) \, \mat{B}^{-1}(t) \, \vec{g}(\vec{s}(t),t)\big\} \,
    \diff t
  \Big]
\end{split}
\end{equation}
\end{proof}

\begin{theorem}[Transformation of SDE Solutions II] \label{the:trans2}
  Assume that processes $\vec{x}_1(t)$, $\vec{x}_2(t)$, $\vec{s}_1(t)$ and
  $\vec{s}_2(t)$ are generated by the stochastic differential
  equations
\begin{align}
    \frac{\diff \vec{x}_1}{\diff t} &= 
      \vec{f}_1(\vec{x}_1,\vec{x}_2,t), 
    \qquad &\vec{x}_1(0) = \vec{x}_{1,0}
   \label{eq:lhsde21a} \\
    \diff \vec{x}_2 &= \vec{f}_2(\vec{x}_1,\vec{x}_2,t) \, \diff t
  + \mat{L}(t) \, \diff \vecbeta, \qquad &\vec{x}_2(0) = \vec{x}_{2,0}
  \label{eq:lhsde21b} \\
    \frac{\diff \vec{s}_1}{\diff t} &= 
      \vec{f}_1(\vec{s}_1,\vec{s}_2,t),
    \qquad &\vec{s}_1(0) = \vec{x}_{1,0}
   \label{eq:lhsde22a} \\
    \diff \vec{s}_2 &= \vec{g}_2(\vec{s}_1,\vec{s}_2,t) \, \diff t
  + \mat{B}(t) \, \diff \vecbeta, \qquad &\vec{s}_2(0) = \vec{x}_{2,0},
   \label{eq:lhsde22b}
\end{align}
where $\mat{L}(t)$ and $\mat{B}(t)$ are invertible matrices for all $t
\ge 0$ and under the measure $\mea{P}$, $\vecbeta(t)$ is a Brownian
motion with diffusion matrix $\mat{Q}(t)$. Then the processes
$\vec{s}_1$ and $\vec{s}_2$ defined as
\begin{align}
  \frac{\diff \vec{s}^*_1}{\diff t} &= 
    \vec{f}_1(\vec{s}^*_1,\vec{s}^*_2,t),
    \qquad &\vec{s}^*_1(0) = \vec{x}_{1,0}
  \label{eq:sstar2a} \\
  \diff \vec{s}^*_2 &= \mat{L}(t) \, \mat{B}^{-1}(t) \, \diff s_2,
  \qquad &\vec{s}^*_2(0) = \vec{x}_{2,0}
  \label{eq:sstar2b}
\end{align}
are weak solutions to the Equations \eqref{eq:lhsde21a} and
\eqref{eq:lhsde21b} under the measure $\tilde{\mea{P}}$ defined by the
relation
\begin{equation}
  \E\left[\frac{\diff \tilde{\mea{P}}}
  {\diff\mea{P}}\,\bigg|\,\alg{F}_t\right] 
  = Z(t).
\label{eq:lh2lhr}
\end{equation}
where
\begin{equation}
\begin{split}
  Z(t) &=
  \exp\Big[
    \int_0^t \big\{\vec{f}_2(\vec{s}^*_1(t),\vec{s}^*_2(t),t)
  - \mat{L}(t) \, \mat{B}^{-1}(t) \,
    \vec{g}_2(\vec{s}_1(t),\vec{s}_2(t),t)\big\}^T \\
  &\qquad \qquad \times \mat{L}^{-T}(t) \,
    \mat{Q}^{-1}(t) \, \diff \vecbeta(t)
  \\
  &\qquad - \frac{1}{2}
    \int_0^t \big\{\vec{f}_2(\vec{s}^*_1(t),\vec{s}^*_2(t),t)
  - \mat{L}(t) \, \mat{B}^{-1}(t) \, \vec{g}_2(\vec{s}_1(t),\vec{s}_2(t),t)\big\}^T \\
  &\qquad \qquad \times \big\{ \mat{L}^T(t) \, \mat{Q}(t) \, \mat{L}(t) \big\}^{-1} \\
  &\qquad \qquad \times \big\{\vec{f}_2(\vec{s}^*_1(t),\vec{s}^*_2(t),t)
  - \mat{L}(t) \, \mat{B}^{-1}(t) \, \vec{g}_2(\vec{s}_1(t),\vec{s}_2(t),t)\big\} \,
    \diff t
  \Big]
\end{split}
\label{eq:lh2z}
\end{equation}
\end{theorem}

\begin{proof}
From equations \eqref{eq:lhsde22a}, \eqref{eq:lhsde22b},
\eqref{eq:sstar2a} and \eqref{eq:sstar2b} we get that
\begin{equation}
\diff \vecbeta(t) = 
  \mat{L}^{-1}(t) \, \diff \vec{s}^*_2 -
  \mat{B}^{-1}(t) \, \vec{g}_2(\vec{s}_1(t),\vec{s}_2(t),t) \, \diff t.
\end{equation}
If we now define
\begin{equation}
  \vectheta(t) = 
    \mat{Q}^{-1}(t) \, \mat{L}^{-1}(t) \, \vec{f}_2(\vec{s}^*_1(t),\vec{s}^*_2(t),t)
  - \mat{Q}^{-1}(t) \, \mat{B}^{-1}(t) \, \vec{g}_2(\vec{s}_1(t),\vec{s}_2(t),t),
\end{equation}
then similarly as in proof of Theorem \ref{the:trans1}, we get
that the process $\tilde{\vecbeta}(t)$ defined as
\begin{equation}
\begin{split}
\diff \tilde{\vecbeta}(t)
&= \diff \vecbeta(t) - \mat{Q}(t) \, \vectheta(t) \diff t \\
&= \mat{L}^{-1}(t) \, \diff \vec{s}^*_2 -
   \mat{B}^{-1}(t) \, \vec{g}_2(\vec{s}_1(t),\vec{s}_2(t),t) \, \diff t \\
&\quad - \mat{Q}(t) \, \mat{Q}^{-1}(t) \,
   \mat{L}^{-1}(t) \, \vec{f}_2(\vec{s}^*_1(t),\vec{s}^*_2(t),t) 
    \, \diff t \\
&\quad + \mat{Q}(t) \, \mat{Q}^{-1}(t) \,
   \mat{B}^{-1}(t) \, \vec{g}_2(\vec{s}_1(t),\vec{s}_2(t),t)
    \, \diff t\\
&= \mat{L}^{-1}(t) \, \diff \vec{s}^*_2
 - \mat{L}^{-1}(t) \, \vec{f}_2(\vec{s}^*_1(t),\vec{s}^*_2(t),t) \, \diff t
\end{split}
\end{equation}
is a Brownian motion with respect to measure $\tilde{\mea{P}}$
and thus $\vec{s}^*_1$ and $\vec{s}^*_2$ are the weak solutions
to the equations \eqref{eq:lhsde21a} and \eqref{eq:lhsde21b}.
\end{proof}

\bibliography{cdpf}

\section{Acknowledgment}

The author would like to thank Aki Vehtari, Jouko Lampinen and Ilkka
Kalliom\"aki for helpful discussions and comments on the manuscript.


\end{document}